\documentclass[prl,showpacs,twocolumn]{revtex4}
\usepackage{epsfig}
\begin{document}

\title{Lateral electron localization
by the induced surface charge}
\author{S. Bednarek\cite{email}, J. Adamowski, and B. Szafran}
\affiliation{Faculty of Physics and Nuclear Techniques,
AGH University of Science and Technology, Krak\'ow, Poland}

\date{\today}

\begin{abstract}
We investigate the problem of the electron interacting with the
charge induced on the metal or dielectric surface. We show that
the interaction between the electron and the induced surface
charge leads to the lateral confinement of the electron. As a
result the electron propagates parallel to the surface not as a
plane wave but as a wave packet of a Gaussian shape.  The electron
moving together with the induced charge can be treated as a new
quasi-particle, which we call inducton. We discuss a possible
experimental evidence for inductons in semiconductor
nanostructures, metal-vacuum, and dielectric-vacuum interfaces.
\end{abstract}


\maketitle

The electron interacting with the charge induced on the metal
(dielectric) surface was a subject of many experimental
\cite{Wein,Gies,Pad,Ge,Miller,Dutt} and theoretical
\cite{Cole,Echen,Faust,Jorg,Gou,Fuhr} papers. In semiconductor
nanostructures, the induced potential leads to a screening
of the electron-electron
interaction and a modification of the potential confining the
charge carriers in the quantum wells, wires, and dots
\cite{Hawr,Hall,Rudin,Bruce,BSA01}. In papers \cite{Cole,Echen}
the classical image-charge potential was applied to study the
electron localization in the direction normal to the metal
surface. The lateral localization of the electron, i.e., parallel
to the metal surface, has been observed by the time-resolved
two-photon photoemission \cite{Pad,Miller}.

In the present Letter we study the electron moving near the
surface of the metal or dielectric. We show that the interaction
with the induced charge leads to the lateral confinement potential
acting on the electron.
In the existing literature,
e.g., \cite{Cole,Echen,Faust,Jorg,Gou,Fuhr}, this
potential has been overlooked due to the neglect of the dependence
of the induced charge distribution on the electron wave function.

We first formulate a two-dimensional (2D) model for a typical
metal-semiconductor planar structure. Next, we extend
the model to the three-dimensional (3D) system and compare the
results of calculations with the available experimental data. We
show that the observed \cite{Pad,Miller} lateral electron
localization can be explained by the effect of the induced surface
charge without a necessity of taking into account neither the
surface corrugation nor the electron-dipol coupling.

We consider a planar structure, in which
an electron is confined in the quasi-2D quantum well
separated from the metal by a blocking barrier
of thickness $d$.  We assume that the quantum well
and the barrier are characterized by the
same dielectric constant $\varepsilon$.
We focus on the properties that
are essentially independent of the specific atomic
and crystal structure of the materials.  Therefore,
we assume the idealized model of the metal, which
consists of a positive background charge
and an arbitrary number of nearly free electrons, and its surface
is the infinite plane.
The single electron
generates the Coulomb field, which induces the positive charge
on the metal surface, which in turn acts on the electron
changing its quantum state.
For our model structure
the interaction of the electron with the induced charge
can be calculated by the image charge method.
We assume that the electron is strongly confined
within the quantum well in the
normal-to-plane (vertical) direction $z$
and the image charge distribution
is a mirror reflection of the electron charge distribution
with respect to the metal surface plane ($z=0$).

The potential energy of the interaction of the electron
with the charge induced on the metal surface can be expressed as
\begin{equation}
W=\frac{\kappa}{2}\int d^3r d^3r^{\prime}
\frac{\varrho_-(\mathbf{r})\varrho_+(\mathbf{r}^{\prime})}
{|\mathbf{r}-\mathbf{r}^{\prime}|} \;,
\label{Hartree}
\end{equation}
where prefactor 1/2 accounts for the self-interaction character of
the electron-image charge interaction and
$\kappa=1/4\pi\varepsilon_0\varepsilon$.
In Eq.~(\ref{Hartree}), there appear the two spatially separated
charge densities: the electron charge density
$\varrho_-(\mathbf{r})=-e|\psi(x,y)|^2\delta(z-d)$
and the image charge density
$\varrho_+(\mathbf{r})=q|\psi(x,y)|^2\delta(z+d)$,
where $e > 0$ is the elementary charge,
$\psi(x,y)$ is the wave function of the electron, and
$q=e$ \cite{q}.

First, in the framework of the simple 2D model,
we perform the variational calculation with
the trial wave function chosen in the form
\begin{equation}
\psi(x,y)=(1/\pi^{1/2}l)\exp[-(x^2+y^2)/2l^2] \;,
\label{psi}
\end{equation}
where variational parameter $l$ determines the radius
of electron localization.
For the Gaussian charge densities
we can apply the effective
interaction potential \cite{eff}
when calculating the potential energy
[Eq.~(\ref{Hartree})].
Expression (\ref{Hartree}) takes on
the form (cf. Eq.~(19) in Ref. \cite{eff})
\begin{equation}
W(l)=-(1/2)(\kappa eq/l)(\pi/2)^{1/2} \;
\mathrm{erfcx}(2^{1/2}d/l)  \;.
\label{W}
\end{equation}
Using wave function (\ref{psi})
we calculate the expectation value of the total energy
$E(l)=\hbar^2/(2m_el^2)+W(l)$,
where $m_e$ is the effective electron mass.
Function $E(l)$, displayed in Fig.\ 1 for $d=a_D$,
possesses the pronounced minimum for finite
radius of electron localization $l=3.61 a_D$
with $E_{min}=-0.158 R_D$,
where $R_D=m_e\kappa^2 e^4/(2\hbar^2)$ is the effective donor rydberg
and $a_D=\hbar^2/(m_e\kappa e^2)$ is the effective Bohr radius.
The results of Fig.\ 1 show the existence of the bound state
with the lateral electron localization.
The electron moves parallel to the interface
not as a plane wave but as a wave packet
of finite spatial extension in
lateral directions.
In order to conveniently picture this motion,
we find it useful to introduce
a new quasi-particle,
which we call \emph{inducton}, since
the confinement potential, which forms this wave packet,
is generated by the induced charge.

\begin{figure}[htbp]{\epsfxsize=70mm
                \epsfbox[22 250 589 781]{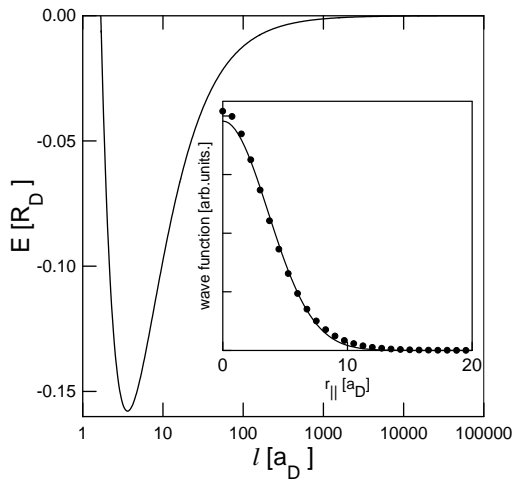}}\newline
\caption{Expectation value $E$ of the ground-state energy of the
inducton calculated with trial wave function (2)
as a function of radius $l$.
Inset: Exact (dots) and variational Gaussian (solid curve,
Eq.~(\ref{psi})) ground-state wave function as a function of
lateral position $r_{\parallel}$. }
\end{figure}

For the stationary states
the Hamiltonian of the inducton can be derived
from the condition $\delta(\langle \psi |T|\psi \rangle
+ W)/\delta \psi^{\star} =0$,
where $T$ is the kinetic energy operator
and $W=W[\psi^{\star},\psi]$ is given by (\ref{Hartree}).
We obtain the Hamiltonian of the form
$H=-\hbar^2\nabla^2/(2m_e)+U(\mathbf{r}_{\parallel},d)$,
where
\begin{equation}
U(\mathbf{r}_{\parallel},d) = - \kappa eq \int
d^2 r^{\prime}_{\parallel}
\frac{|\psi(\mathbf{r^{\prime}_{\parallel}})|^2}
{[(\mathbf{r}_{\parallel}-\mathbf{r}^{\prime}_{\parallel})^2+4d^2]^{1/2}}
\label{U}
\end{equation}
and $\mathbf{r}_{\parallel}=(x,y)$.
Let us discuss the physical content of Hamiltonian $H$.
Quantity $U$ [Eq.~(\ref{U})]
is the Hartree potential energy due to the charge induced
on the metal (dielectric) surface,
but is not the potential energy of the electron,
which absolute value is two times smaller  [cf. Eq.~(\ref{Hartree})].
As a result, eigenfunctions $\psi_n$ of Hamiltonian
$H$ can be identified with the electron wave functions
in the inducton states,
but the eigenvalues of $H$, $E_n$, do not determine the total
energy of the system.  Instead $E_n$
possess the same interpretation
as the one-electron energies in the
self-consistent method, e.g., Hartree-Fock method.
We note that one can obtain the same Hartree potential (\ref{U})
using the quantum linear response theory.

\begin{figure}[htbp]{\epsfxsize=70mm
                \epsfbox[14 218 581 754]{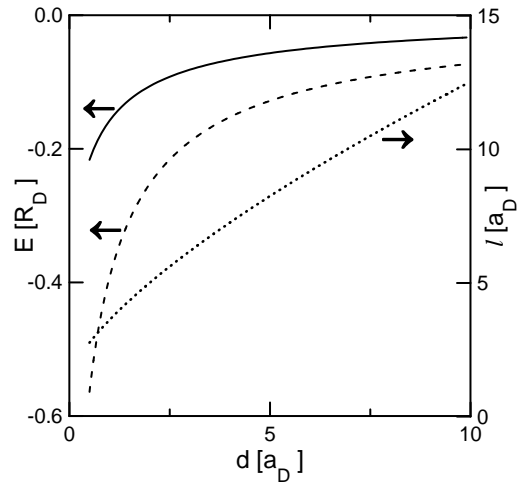}}\newline
\caption{Ground-state energy: total (solid curve) and one-electron
(dashed curve), and radius $l$ (dotted curve) of the inducton as
functions of distance $d$ from the quantum well to metal surface.}
\end{figure}

Due to the interaction with the induced charge the electron
is not an isolated particle.
During the transition
between quantum states the induced charge
can either transfer its energy to the electron
(the transition with full relaxation)
or release it partly or fully into the metal
(the transitions with partial relaxation
or without relaxation).
In the first case, we can observe the total energy of the inducton,
while in the last case, we can measure
the one-particle energy,
similarly as in the many-electron
system described by the Hartree-Fock method
under condition that the Koopmans theorem is satisfied.
Figure 2 displays the one-electron lowest energy level,
total energy, and radius $l$ of the inducton.
We note that the inducton radius is comparable with $d$.

The results depicted in Figs.\ 1 and 2
have been obtained with Gaussian trial
wave function (\ref{psi}).
In order to check the
validity of our choice of wave function (\ref{psi}),
we have solved the eigenequation of $H$
by the numerical self-consistent method \cite{itime}
on the one-dimensional radial mesh.
In the present paper, we limit our attention
to the $s$-like solutions with the rotational
symmetry.
The numerical procedure yields the results,
which can be treated as ''exact''.  We have found
(cf. inset of Fig.\ 1) that the ground-state wave function
obtained by the self-consistent procedure
is almost the same as the optimum
trial wave function [Eq.~(2)], which means that
the exact ground-state wave function
can be very well approximated by the Gaussian of form (\ref{psi}).
This proves the consistency of our simple model
based on Eq.~(\ref{psi}).

The electron motion parallel to the interface can be described by
a plane wave, but then the induced charge density is zero, total
energy $E \geq 0$, and the inducton states are not formed. In the
inducton states with $E<0$, in which the electron wave function is
localized, potential $U$ [Eq.~(\ref{U})] does not vanish and is
attractive. In these states, the momentum of the electron is not
defined. The inducton can move in the lateral direction with the
non-zero momentum but the electron wave function in the inducton
state is a wave packet.

The inductons considered so far were formed in the quantum well;
so, they were described as the 2D systems. Let us generalize our
approach to the 3D systems. For this purpose we abandon the
assumption of the quantum-well confinement and consider the
electron moving in the half-space outside the blocking barrier.
The electron induces the charge on the metal surface and due to
the interaction with the induced charge the 3D wave packet is
formed. The system is described by the corresponding 3D
Hamiltonian, which has the form of its 2D counterpart with
potential energy $U(\mathbf{r})=U(\mathbf{r}_{\parallel},z)$
resulting from the charge induced on the metal surface and
potential of the blocking barrier, i.e., $U(\mathbf{r}) = \infty$
for $0 < z \leq d$, and $U(\mathbf{r})$ is given by the 3D version
of formula (\ref{U}) for $z > d$. The image charge density is the
mirror reflection of the electron charge density in the $z=0$
plane, i.e., $\varrho_{+,n}(\mathbf{r})=q|\psi_n(x,y,-z)|^2$.

\begin{figure}[htbp]{\epsfxsize=75mm
                \epsfbox[45 486 531 750]{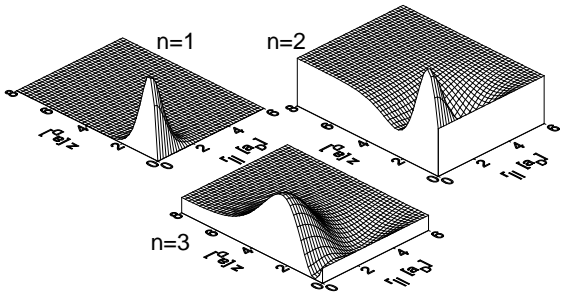}}\newline
\caption{Wave functions of the three lowest-energy $s$-like states
plotted as functions of cylindrical coordinates $r_{\parallel}$
and $z$. }
\end{figure}

Assuming the cylindrical symmetry,
we solve the corresponding 3D eigenproblem
on the 2D mesh $(r_{\parallel},z)$
by the self-consistent procedure.
Figure 3 depicts the $s$-like wave functions, which correspond
to the lowest-energy levels.
The ground-state wave function ($n=1$), as a function of $r_{\parallel}$,
has the Gaussian shape.
The excited-state wave functions ($n=2,3$) clearly exhibit
the non-separability of variables $r_{\parallel}$ and $z$.

Figure 4 shows that with the increasing blocking barrier
thickness the energy levels shift upwards
and the interlevel separations decrease.
This feature is in a qualitative agreement with
the experimental data \cite{Pad}.
We note that the energy levels do not form
the hydrogenlike Rydberg series.

We expect that the quasi-2D inductons can occur in planar
semiconductor nanostructures, which contain
the quantum-well layer and are covered by
the metal layer.  The largest binding energy
of the inducton
should be observed in semiconductors
with small dielectric constant
and large electron band mass.  Due to the larger
valence band mass, the hole inducton should
be more stable than the electron inducton.

The 3D inductons can appear in semiconductor structures
of the type of the field effect transistor
with the isolated gate.  Due to the increase
of the binding energy with the decreasing blocking
barrier thickness, the inductons formed in the structures
containing the thin blocking barrier will be the most
strongly bound.  A possible experimental evidence
of such inducton states is within the reach of modern
experimental setups.

\begin{figure}[htbp]{\epsfxsize=75mm
                \epsfbox[14 332 583 815]{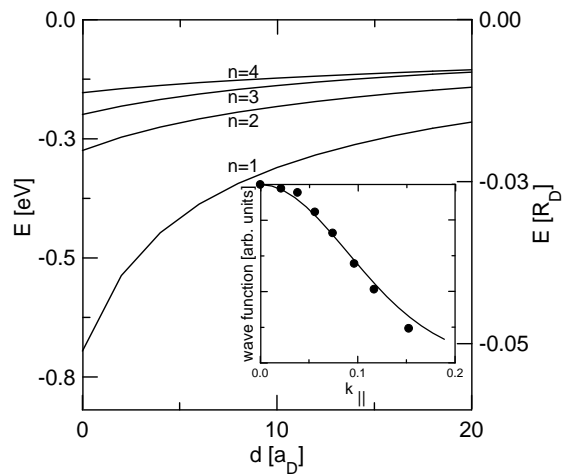}}\newline
\caption{One-electron $s$-like energy levels calculated with
$\varepsilon=1$ as functions of thickness $d$ of the blocking
layer. Inset: Fourier transform of the ground-state inducton wave
function (solid curve) vs $k_{\parallel}$ (in \AA$^{-1}$). Dots
correspond to the experimental data taken from Fig.\ 4(A) of Ref.
\cite{Miller} for time delay 266 fs.}
\end{figure}

However, the so-called image-potential electron states \cite{Jorg}
were observed \cite{Pad,Gies,Miller} near the metal-vacuum
interface. The theoretical interpretation of these states is based
upon the notion of Rydberg states \cite{Cole,Echen}. In this
interpretation \cite{Cole,Echen}, the potential energy of the
interaction of the electron with its image charge is assumed in
the form $U_{image}(z) = -\kappa eq/(4z)$, which is valid for the
classical point charge. In quantum mechanics, the charge density
of the electron is determined by a wave function. The image of the
spread-out charge is also spread out. The potential energy of the
electron should be calculated according to Eq.~(\ref{Hartree}) and
the induced potential should depend on lateral position
$r_{\parallel}$ [cf. Eq.~(\ref{U})]. Potential $U_{image}$, used
in Refs. \cite{Cole,Echen}, corresponds to either the classical
particle or the quantum particle with the lateral wave function
given by a Dirac delta function. In spite of this the authors
\cite{Cole,Echen} use the plane wave for the lateral motion of the
electron, which is a serious inconsistency. If the electron
travels in the $x-y$ plane as the plane wave (cf. Fig.\ 1 for $l
\longrightarrow \infty$), its image charge is completely
delocalized and the induced potential is zero \cite{resp}. This
inconsistency is repeated in all the other papers on this subject,
e.g., \cite{Jorg,Faust,Fuhr}.

Based on the results of the present paper, we propose an
alternative interpretation of the states observed in Refs.
\cite{Pad,Miller}. According to our proposition these findings
\cite{Pad,Miller} can be treated as the experimental evidence for
the formation of the inducton. To support this interpretation we
have calculated the Fourier transform of the ground-state wave
function of the inducton. The results (inset of Fig.\ 4) very well
agree with the experimental data \cite{Miller}. Using the results
of Fig.\ 4 for $d=0$, we have calculated the energy differences
$\Delta_{mn}=E_m-E_n$ between the one-particle energy levels and
compared them with the experimental data \cite{Pad}. The
calculated (measured \cite{Pad}) energy differences are
$\Delta_{21}$ = 0.41 (0.43) eV, $\Delta_{31}$ = 0.49 (0.50) eV,
and $\Delta_{41}$ = 0.54 (0.56) eV (energy separation
$\Delta_{21}$ estimated from Ref. \cite{Miller} is 0.45 eV). We
see that the agreement with experiment is very good. The fact that
the one-particle energy levels are observed in experiments
\cite{Pad,Miller} can result from the slow relaxation of the
induced charge during the photoemission. We note that the present
model is valid if the electron does not penetrate the metal
\cite{Roul}, i.e., if the metal surface is covered by a thin
dielectric layer.  This condition is satisfied in the experiments
\cite{Pad,Miller}.

In summary, we have shown that the charge induced on the metal
(dielectric) surface by the nearby electron creates the lateral
confinement potential, which in turn leads to the lateral
localization of the electron. The collective
electron-induced-charge motion can be conveniently described with
the help of the new quasi-particle, the inducton. According to the
reinterpretation we have proposed the localized states observed
near the metal surface are the inducton states. The electron
described by the plane wave does not generate any induced charge
potential. The plane-wave states can only be created if the
excitation energy is larger than the continuum-energy threshold.
The electron localization in the vertical direction is only
possible if the lateral electron localization occurs. Electrons
states which are not laterally localized cannot be bound to the
metal surface.

We expect the inducton states to be universal
for the charge carriers moving near the metal or dielectric surfaces
and should observed at the metal-vacuum and dielectric-vacuum
interfaces as well as in semiconductor nanostructures.

\acknowledgments
This work has been supported in part by the Polish Government
Scientific Research Committee (KBN).


\begin{thebibliography}{99}
\bibitem{q}
The present theory is also valid when replacing
the metal by the dielectric with dielectric
constant $\varepsilon^{\prime}$, where
$\varepsilon^{\prime} > \varepsilon$.
In this case
$q=e(\varepsilon^{\prime}-\varepsilon)/
(\varepsilon^{\prime}+\varepsilon)$.
\end{thebibliography}
\end{document}